\pdfoutput=1
\documentclass[11pt]{article}
\usepackage{colortbl}
\usepackage{array}
\usepackage{booktabs} 
\usepackage{amsmath}
\usepackage{amssymb}

\usepackage[final]{acl}
\usepackage{multirow}
\usepackage{multicol}
\usepackage{times}
\usepackage{latexsym}
\usepackage{subfig}
\usepackage{array} % 允许使用m{}列类型
\usepackage{graphicx} % 允许多行内容自动换行

\usepackage[T1]{fontenc}
\usepackage[utf8]{inputenc}

\usepackage{microtype}

\usepackage{inconsolata}

\usepackage{graphicx}

\newcommand{\ourmethod}{CoVE}
\newcommand{\Mod}[1]{\ \mathrm{mod}\ #1}

\definecolor{darkgreen}{RGB}{0,100,0}

\title{\ourmethod: Compressed Vocabulary Expansion Makes Better LLM-based Recommender Systems}

\author{
 \textbf{Haochen Zhang\textsuperscript{1}},
 \textbf{Tianyi Zhang\textsuperscript{1}},
 \textbf{Junze Yin\textsuperscript{1}},
 \textbf{Oren Gal\textsuperscript{2}},
\\
 \textbf{Anshumali Shrivastava\textsuperscript{1}},
 \textbf{Vladimir Braverman\textsuperscript{1,3}}
\\
\\
 \textsuperscript{1}Rice University,
 \textsuperscript{2}Swarm \& AI, University of Haifa,
 \textsuperscript{3}John Hopkins University
\\
 \small{
   \textbf{Correspondence:} \href{hz112@rice.edu}{hz112@rice.edu}
 }
}

\begin{document}
\maketitle
\begin{abstract}
Recommender systems play a pivotal role in providing relevant content to users. With the rapid development of large language models (LLMs), researchers have begun utilizing LLMs to build more powerful recommender systems. However, existing approaches that focus on aligning LLMs with recommendation tasks do not fully leverage their sequential information processing capabilities, leading to suboptimal performance. 

In this paper, we propose a novel system called \emph{\underline{co}mpressed \underline{v}ocabulary \underline{e}xpansion (\ourmethod)}. In \ourmethod, each item is assigned a unique ID within the expanded vocabulary. Our framework effectively capitalizes on sequence understanding abilities of LLMs, significantly enhancing their performance on recommendation tasks. Additionally, we compress the embedding layer, making \ourmethod\ practical for large-scale industrial applications. The effectiveness and performance of \ourmethod\ are demonstrated through comprehensive experiments on multiple recommendation datasets and comparisons with prior works. Our code can be found at \url{https://github.com/HaochenZhang717/CoVE-official-Repo}.
\end{abstract}

\section{Introduction}

\begin{figure*}[!ht]
    \centering
  \includegraphics[width=\linewidth]{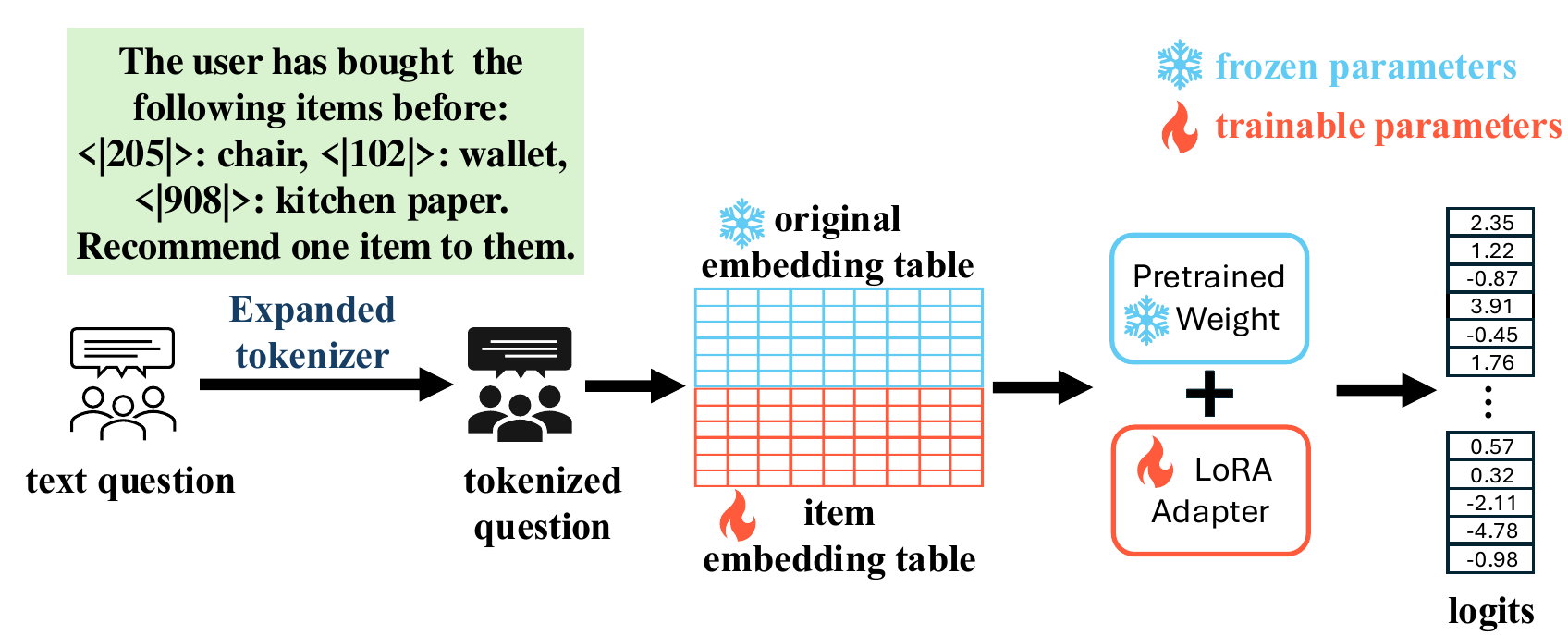} 
  \caption{An overview of our \ourmethod\ framework. Each item is assigned a unique ID in the tokenizer. It is then mapped to a distinct embedding. After that, our framework finetunes item embeddings, transformer weights, and the lm\_head to align LLMs with sequential recommendation tasks.}
\label{figure-1}
\end{figure*}

Large language models (LLMs), like GPT series \cite{gpt1,gpt2,gpt3,gpt4}, LLaMA \cite{llama1,llama2}, and Deepseek \cite{deepseek}, have demonstrated strong capabilities for reasoning and problem-solving \cite{achiam2023gpt, touvron2023llama, almazrouei2023falcon}. LLMs are typically pretrained on a large corpus and then finetuned on smaller datasets for downstream tasks \cite{xiong2024iterative, rafailov2024direct, wei2021finetuned}. The pretrain-and-finetune paradigm achieves the state-of-the-art performance in many tasks \cite{w24,hxy+22}. 

On the other hand, in current big data era, people increasingly face challenges related to information overload \cite{boka2024survey}. This brings sequential recommendation problems, where a recommender system analyzes a user's historical interaction data to predict the next item that is likely to interest them \cite{bzk+24}. With the rise of LLMs, a large body of research \cite{hzl+23,lww+24,hxz+24,wsz24,bzk+24,yyy+24} has been dedicated to developing proper frameworks to effectively leverage LLMs for sequential recommender systems.

Among existing works, there are two main ways to utilize LLMs for recommender systems. One approach is to use embeddings provided by LLMs to initialize non-LLM recommender systems. For example, \citet{yuan2023go,bzk+24} show that LLM embeddings improve the performance of several non-LLM deep learning sequential recommender models \cite{sun2019bert4rec, sasrec, gru4rec}. Despite the positive results from this pipeline, using LLMs solely as embedding providers takes advantage of their embedding ability but does not effectively leverage their content comprehension capability. This can potentially result in suboptimal performance.

The other approach is finetuning LLMs to directly provide recommendations \cite{bao2023bi, bzk+24} which use the pipeline consisting of 1) finetuning LLMs on a recommendation dataset, 2) using finetuned LLMs to generate the next item's title, and 3) conducting embedding retrieval to avoid hallucinated item titles. \citet{bao2023bi} shows that this pipeline produces more accurate recommendations than prior works: it leverages LLMs' token prediction ability, which may address the limitations of the first line of work. However, there are issues in the design of this framework. 1) Under this framework, LLM needs to accurately predict the next item's title, which consists of multiple tokens, and this is a difficult task. 2) The recommendation hallucination problem is inevitable because the generated item's title may not exist in the item space. Though embedding retrieval can be a remedy, this solution is far from ideal. 3) Generating text with LLM is very time-consuming.

As an alternative to the two broad lines of LLM-based recommendation, we propose a novel system herein called \emph{\underline{co}mpressed \underline{v}ocabulary \underline{e}xpansion (\ourmethod)} which is significantly different from the existing pipeline. \ourmethod\ first expands the tokenizer's vocabulary and embedding table by assigning each item in the item space a unique token in the tokenizer and a corresponding unique embedding in the embedding table. During finetuning, the item embedding table is updated together with the LoRA adapter \cite{hu2021lora} to align the pretrained LLM with the recommendation task, as shown in Figure~\ref{figure-1}. During inference, \ourmethod\ can provide a recommendation by generating one token representing the unique ID of the next item, instead of generating the full title. The advantage of \ourmethod\ over existing frameworks is three-fold: 
\begin{enumerate}
    \item The expanded vocabulary enables \ourmethod\ to directly leverage the next-token prediction capability of the LLM, eliminating any chance of hallucination.

    \item \ourmethod\ naturally enables leveraging temporal correlations by providing the entire context to the LLM as a sequence and utilizing next-token prediction.  

    \item During inference, \ourmethod\ operates in a non-generative manner, outputting logits and making recommendations based on them. This significantly speeds up inference compared to the aforementioned framework.
\end{enumerate}

The inspiration for \ourmethod\ comes from vocabulary expansion for domain-specific LLM alignment \cite{cui2023efficient, wang2023huatuo, liu2024gold}, where it has been shown that expanding the vocabulary in a downstream domain improves an LLM’s ability in content comprehension and generation. However, in \ourmethod\ the memory cost of an additional embedding table is prohibitive in large-scale industrial scenarios, where the item space may contain millions of products. For instance, The newest version of the Amazon Review dataset contains 48.19 million items in total\footnote{\url{https://amazon-reviews-2023.github.io}.}.
When finetuning on such a large dataset, the item embedding table itself takes up approximately 96GB in GPU memory, which makes training infeasible for even the most powerful GPUs. Therefore, a key research question that we need to address in this paper is:
\begin{center} 
    {\it How can we make \ourmethod\ memory-efficient while preserving its superior performance compared to the state-of-the-art results?} 
\end{center}

The solution to this problem is using embedding layer compression methods which have been explored by prior works \cite{desai2022random, 16, 47} to alleviate the memory problem brought by large embedding tables. Inspired by them, we propose to use a hash function to compress the item embedding table to make \ourmethod \ more memory efficient. In summary, the contribution of this study is as follows:
\begin{itemize}
    \item We propose a novel finetuning framework, \ourmethod, to align pretrained LLMs with sequential recommendation tasks. \ourmethod\ fully leverages LLMs' sequential information processing capabilities, achieving up to a 62\% improvement in recommendation accuracy over existing methods.
    \item \ourmethod~achieves approximately 100 times faster inference speed than the existing finetune-and-retrieval framework due to the expanded vocabulary.
    \item Embedding layer compression ensures the memory efficiency of \ourmethod. Experiments show that we can compress the embedding table by up to 16 times while still achieving better results than prior methods. 
\end{itemize}

\section{Proposed Framework}
\begin{table*}[!ht]
  \centering
  \begin{tabular}{|p{3cm}|p{10cm}|}
    \hline
    \textbf{Task Instruction:} & Given a list of video games the user has played before, please recommend a new video game that the user likes to the user. \\
    \hline
    \textbf{Task Input:} & 
    Here is the list of video games that the user has played before:
    <|16659|>: \texttt{"Homeworld - PC"}, <|5641|>: \texttt{"Starcraft"}, <|4375|>: \texttt{"StarCraft Expansion Pack: Brood War - PC"}, <|14747|>: \texttt{"Sid Meier's Alpha Centauri - PC"}
    \\
    \hline
    \textbf{Task Output:} & <|16673|>: \texttt{"Total Annihilation Gold (Mac)"} \\
    \hline
  \end{tabular}
  \caption{A tuning sample for the recommendation system. <|$\cdot$|> denotes the unique ID for each item in item space.}
  \label{tab:input-output-form}
\end{table*}

\paragraph{Finetuning Task Formulation}

Our goal is to leverage an LLM to construct a recommender system. To achieve this, we finetune the LLM on a recommendation dataset. The key difference between our finetuning task and common finetuning tasks is that we also tune the embedding table and lm\_head. Table~\ref{tab:input-output-form} provides a sample from the finetuning dataset.
During training, the Task Instruction, Task Input, and Task Output are fed into the LLM, which is trained to minimize the next-token prediction loss. At inference, we provide only the Task Instruction and Task Input to the LLM and expect the first token it generates to be the unique ID of the predicted item.
In this way, \ourmethod\ does not require the LLM to generate the complete name of an item as in \citet{bao2023bi}. Instead, obtaining the output of lm\_head, i.e., logits, is sufficient for making predictions. The logits form a vector whose dimensionality equals the sum of the number of tokens in the original vocabulary and the cardinality of the item space, denoted as $|\mathcal{I}|$. At inference, we only need to consider the last $|\mathcal{I}|$ entries of the logits for item prediction.

\paragraph{\ourmethod}

As shown in Figure~\ref{figure-1}, \ourmethod\ involves tokenizer vocabulary expansion, item embedding table expansion, and a low-rank weight adapter. The core idea of \ourmethod\ is to assign each item a unique embedding. To achieve this, we expand the tokenizer's vocabulary by adding item IDs, such as <|205|>, <|102|>, and <|908|> in Figure~\ref{figure-1}. These item IDs are then encoded as distinct embeddings. \ourmethod\ finetunes item embeddings, weights in transformer blocks, and weights in lm\_head to align pretrained LLMs with sequential recommendation tasks.

\paragraph{Hashing-based Embedding Compression}

Due to the large item space, item embeddings must be compressed to enable efficient GPU training. We propose a hashing-based embedding compression technique, inspired by \citet{compositionalembeddings}. Let $\mathcal{I}$ denote the item space and $\mathcal{S}$ the latent shared item space, where $|\mathcal{S}| \ll |\mathcal{I}|$. We define $\mathcal{H}$ as a universal hash family, where each hash function $h: \mathcal{I} \to \mathcal{S}$ maps items to the shared item space. We sample $k$ such hash functions, denoted as $h_1, \dots, h_k$. Each shared item $s \in \mathcal{S}$ is assigned a high-dimensional embedding, and an item's embedding is obtained by averaging the embeddings of its hashed counterparts in the shared space. Specifically, the embedding for an item $i$ is given by
\begin{align*}
    \frac{1}{k} \sum_{j=1}^{k} \mathbf{e}_{h_j(i)},
\end{align*}
where $\mathbf{e}_s$ represents the high-dimensional embedding of the latent shared item $s$.

We select a universal hash function based on simple arithmetic operations--addition, multiplication, and modulo--which can be efficiently computed on GPUs. Specifically, given an item's index $i$, the hash code for its shared embedding is computed as
\begin{align*}
    h(i) = ((a i + b) \Mod p) \Mod |\mathcal{S}|,
\end{align*}
where $p$ is a large prime number, and $a \in \{1, \dots, p\}$ and $b \in \{0, \dots, p\}$ are randomly chosen parameters.

\section{Experiments}

\subsection{Experiment Settings}
\label{sec:settings}

\paragraph{Datasets}

We conduct experiments on four public real-world benchmarks: Amazon Video Games \cite{McAuley_AmazonData}, Amazon Beauty \cite{hm16}, Amazon Toys \& Games \cite{hm16}, and Amazon Sports \& Outdoors \cite{hm16}. The statistics of these datasets are shown in Table~\ref{tab:datasets}. For Amazon Beauty, Amazon Toys \& Games, and Amazon Sports \& Outdoors, we compare \ourmethod\ with the results reported in \cite{tiger}; therefore, we adopt the same train-test split as in \cite{tiger}. For Amazon Video Games, we compare \ourmethod\ with the results reported in \cite{bao2023bi}; hence, we adopt the same train-test split as in \cite{bao2023bi}.

\paragraph{Architecture, Hyperparameters, and Implementation}

For experiments on Amazon Beauty, Toys \& Games, and Sports \& Outdoors, we use LLaMA-3.2-3B \cite{dubey2024llama} as the backbone model and finetune it for up to 10 epochs. We set the learning rate to $10^{-4}$, the batch size to 32, the LoRA rank to 8, and the LoRA alpha to 16.
For experiments on Amazon Video Games, we maintain the same hyperparameters but change our backbone model to LLaMA-2-7B \cite{llama2} and apply 4-bit QLoRA for a fair comparison with BIGRec \cite{bao2023bi}. All experiments are conducted on an NVIDIA A100 GPU.

\begin{table}[!ht]
  \centering
  \begin{tabular}{lcccc}
    \toprule
    \textbf{Datasets} & \textbf{\# items} & \textbf{\# sequences} \\
    \toprule
    Video Games & 17,408 & 149,796 \\
    Beauty & 12,101 & 22,363\\
    Toys \& Games & 11,924 & 19,412\\
    Sports \& Outdoors & 18,357 & 35,598\\
    \toprule
  \end{tabular}
  \caption{Statistics of datasets.}
  \label{tab:datasets}
\end{table}

\paragraph{Baselines}

To verify the efficacy of the proposed framework, we compare it with various prior methods, including GRU4Rec \cite{gru4rec}, BERT4Rec \cite{sun2019bert4rec}, Caser \cite{caser}, SASRec \cite{sasrec}, P5 \cite{p5}, HGN \cite{hgn}, S$^3$-Rec \cite{s3rec}, FDSA \cite{fdsa}, TIGER \cite{tiger}, DROS \cite{dros}, GPT4Rec \cite{li2023gpt4rec}, and BIGRec \cite{bao2023bi}. A brief description of these methods is provided in Appendix~\ref{app:prior method}. BIGRec and \ourmethod\ finetune LLMs to build recommender systems, whereas BIGRec adopts a finetune-and-retrieval framework. The rest of the above work do not directly finetune LLMs.

\paragraph{Metrics}

Like many existing works, we adopt normalized discounted cumulative gain at rank $K$ (NG@K, or NGK) \cite{wwl+13,gshp20,jc15} and hit rate at rank $K$ (HR@K, or HRK) \cite{ez23}. These metrics help quantify how effectively a system ranks relevant items for a given user. 
Specifically, NG@K measures the ranking quality of recommended items by considering both the relevance of items and their positions in the ranked list and is defined as: 
\begin{align*}
    NG@K = \frac{\sum_{i=1}^{K} \frac{\mathrm{rel}_i}{\log_2(i+1)}}{\sum_{i=1}^{K} \frac{\mathrm{rel}_i^{*}}{\log_2(i+1)}},
\end{align*}
where $\mathrm{rel}_i$ is the relevance score of the item at position $i$ and $\mathrm{rel}_i^{*}$ represents the relevance scores of the items sorted in optimal order.
Because of this construction, we can see that $NG@K \in [0, 1]$, where 1 denotes the best possible ranking.

HR@K, on the other hand, measures the percentage of users for whom at least one relevant item appears in the top $K$ recommendations and is defined as:
\begin{align*}
    HR@K = \frac{\sum_{u=1}^{U} \mathbb{I}(\exists i \in \{1, \dots, K\}, \mathrm{rel}_{u,i} > 0)}{U},
\end{align*}
where $U$ is the total number of users, $\mathbb{I}(\cdot)$ is an indicator function, and $\mathrm{rel}_{u,i}$ is the relevance of item $i$ for user $u$. 
In our experiments, we have shown that our pipeline \ourmethod\ results in higher NG@K and HR@K for various choices of $K$.

\subsection{Main Results}

\begin{figure*}[!ht]
\subfloat[The percentage improvement of \ourmethod\ compared with the state-of-the-art results in NG@5, NG@10, HR@5, and HR@10.]{
  \includegraphics[width= 0.45\textwidth]{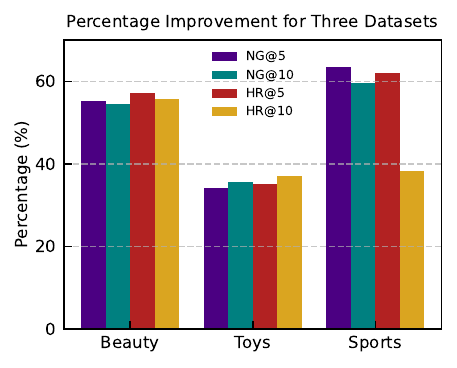}}
  \hfill
  \subfloat[{The percentage improvement of \ourmethod\ compared to BigRec on the Video Games dataset. BigRec uses finetune-and-retrieval framework.}]{
  \includegraphics[width= 0.49\textwidth]{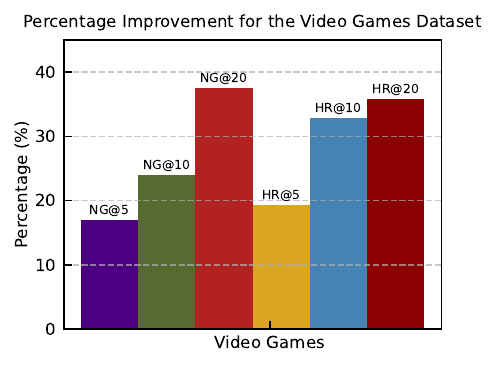}}
  \caption{The percentage improvement of \ourmethod\ over the best metric achieved by TIGER~\cite{tiger} (left) BIGRec~\cite{bao2023bi} (right) on four different datasets: ``Beauty'', ``Toys'', ``Sports'', and ``Video Games''. The metric values for \ourmethod\ and prior works on these datasets are presented in Tables~\ref{tab:compact_table} and~\ref{tab:first_main_table}.}
  \label{fig:3}
\end{figure*}

\begin{table*}[!ht]
  \centering
  \renewcommand{\arraystretch}{1.2} % Increases row height for better spacing
  \setlength{\tabcolsep}{3pt}       % Reduces column spacing
  
  \begin{tabular}{p{1.1cm} p{1.1cm} p{1.1cm} p{1.1cm} p{1.1cm} p{1.1cm} p{1.1cm} p{1.1cm} p{1.1cm} p{1.1cm} p{1.1cm} p{1.1cm}}
    \toprule
    \rowcolor{gray!20} % Light gray header row
    \multirow{2}{*}{\textbf{Datasets}} & \multirow{2}{*}{\textbf{Metric}} & \textbf{GRU 4 Rec} & \textbf{BERT 4 Rec} & \multirow{2}{*}{\textbf{Caser}} & \textbf{SAS Rec} & \multirow{2}{*}{\textbf{P5}} & \multirow{2}{*}{\textbf{HGN}} & \textbf{S$^3$-Rec} & \multirow{2}{*}{\textbf{FDSA}} & \multirow{2}{*}{\textbf{TIGER}} & \multirow{2}{*}{{\bf \ourmethod}} \\
    \toprule

    \multirow{4}{*}{\textbf{Beauty}}  
    & \textbf{NG5}   & 0.0099 & 0.0124 & 0.0131 & 0.0249 & 0.0107 & 0.0206 & 0.0244 & 0.0163 & \underline{0.0321} & \cellcolor{darkgreen!10} {\bf 0.0498}\\ 
    & \textbf{NG10}  & 0.0137 & 0.0170 & 0.0176 & 0.0318 & 0.0136 & 0.0266 & 0.0327 & 0.0208 & \underline{0.0384} & \cellcolor{darkgreen!10} {\bf 0.0593}\\ 
    & \textbf{HR5}   & 0.0164 & 0.0203 & 0.0205 & 0.0387 & 0.0163 & 0.0325 & 0.0387 & 0.0267 & \underline{0.0454} & \cellcolor{darkgreen!10} {\bf 0.0714}\\ 
    & \textbf{HR10}  & 0.0283 & 0.0347 & 0.0347 & 0.0605 & 0.0254 & 0.0512 & 0.0647 & 0.0407 & \underline{0.0648} & \cellcolor{darkgreen!10} {\bf 0.1009}\\ 
    \midrule
    
    % Toys Dataset
    % \rowcolor{blue!10} % Light blue for dataset name row
    \multirow{4}{*}{\textbf{Toys}}  
    & \textbf{NG5}   & 0.0059 & 0.0071 & 0.0107 & 0.0306 & 0.0050 & 0.0221 & 0.0294 & 0.0140 & \underline{0.0371} & \cellcolor{darkgreen!10} {\bf 0.0509}\\ 
    & \textbf{NG10}  & 0.0084 & 0.0099 & 0.0141 & 0.0374 & 0.0066 & 0.0277 & 0.0376 & 0.0189 & \underline{0.0432} & \cellcolor{darkgreen!10} {\bf 0.0595}\\ 
    & \textbf{HR5}   & 0.0097 & 0.0116 & 0.0166 & 0.0463 & 0.0070 & 0.0321 & 0.0443 & 0.0228 & \underline{0.0521} & \cellcolor{darkgreen!10} {\bf 0.0719}\\ 
    & \textbf{HR10}  & 0.0176 & 0.0203 & 0.0270 & 0.0675 & 0.0121 & 0.0497 & 0.0700 & 0.0381 & \underline{0.0712} & \cellcolor{darkgreen!10} {\bf 0.0986}\\ 
    \midrule

    % Sports Dataset
    % \rowcolor{blue!10} % Light blue for dataset name row
    \multirow{4}{*}{\textbf{Sports}}  
    & \textbf{NG5}   & 0.0086 & 0.0075 & 0.0072 & 0.0192 & 0.0041 & 0.0120 & \underline{0.0204} & 0.0156 & 0.0181 & \cellcolor{darkgreen!10} {\bf 0.0296}\\ 
    & \textbf{NG10}  & 0.0110 & 0.0099 & 0.0097 & \underline{0.0249} & 0.0052 & 0.0159 & 0.0240 & 0.0156 & 0.0225 & \cellcolor{darkgreen!10} {\bf 0.0359}\\ 
    & \textbf{HR5}   & 0.0129 & 0.0115 & 0.0116 & 0.0233 & 0.0061 & 0.0189 & 0.0251 & 0.0182 & \underline{0.0264} & \cellcolor{darkgreen!10} {\bf 0.0428}\\ 
    & \textbf{HR10}  & 0.0204 & 0.0191 & 0.0194 & 0.0350 & 0.0095 & 0.0313 & 0.0385 & 0.0288 & \underline{0.0400} & \cellcolor{darkgreen!10} {\bf 0.0624}\\ 
    \bottomrule
  \end{tabular}
  
  \caption{Performance comparison of different recommendation models across three datasets: Beauty, Toys and Games, and Sports and Outdoors. The results of \ourmethod\ shown in the table use a compression rate of 2. The best metrics are always achieved by using \ourmethod\ and are all in bold font in this table, and the best metric achieved by prior works (see the {\bf Baselines} paragraph in Section~\ref{sec:settings}) is underlined.}  
  \label{tab:compact_table}
\end{table*}

\begin{table*}[!ht]
  \centering
  \begin{tabular}{llccccccccccc}
    \toprule
    \rowcolor{gray!20}
    \textbf{Datasets} & \textbf{Model} & {\bf NG5} & {\bf NG10} & {\bf NG20} & {\bf HR5} & {\bf HR10} & {\bf HR20} \\
    \toprule
    \multirow{8}{*}{Video Games}     
    & GRU4Rec & 0.0018 & 0.0024 & 0.0030 &0.0024 & 0.0041 & 0.0069 \\
    & Caser & 0.0019 & 0.0024 & 0.0035 & 0.0032 & 0.0048 & 0.0092 \\
    & SASRec & 0.0015 & 0.0024 & 0.0035 & 0.0021 & 0.0037 & 0.0057\\
    & P5 & 0.0007 & 0.0010 & 0.0017 & 0.0012 & 0.0023 & 0.0049\\
    & DROS & 0.0013 & 0.0016 & 0.0022 & 0.0019 & 0.0027 & 0.0052\\
    & GPT4Rec-LLaMA & 0.0000 & 0.0001 & 0.0001 & 0.0000 & 0.0002 & 0.0002 \\
    & BIGRec & \underline{0.0189} & \underline{0.0216} & \underline{0.0248} & \underline{0.0243} & \underline{0.0329} & \underline{0.0457} \\
    & \cellcolor{darkgreen!10} \ourmethod & \cellcolor{darkgreen!10} {\bf 0.0221} & \cellcolor{darkgreen!10} {\bf 0.0268} & \cellcolor{darkgreen!10} {\bf 0.0341} & \cellcolor{darkgreen!10} {\bf 0.0290} & \cellcolor{darkgreen!10} {\bf 0.0437} & \cellcolor{darkgreen!10} {\bf 0.0621} \\
    \toprule
  \end{tabular}
  \caption{Performance comparison of different recommender systems on Video Game dataset. The best metrics are always achieved by using \ourmethod\ and are all in bold font in this table, while the best metric achieved by prior works (see the {\bf Baselines} paragraph in Section~\ref{sec:settings}) is underlined.}
  \label{tab:first_main_table}
\end{table*}

\paragraph{Recommendation Accuracy} Table~\ref{tab:compact_table} compares \ourmethod\ with prior works on the Beauty, Toys \& Games, and Sports \& Outdoors datasets. The results show that \ourmethod\ provides recommendations more consistent with users' interests. The percentage improvement of \ourmethod\ over prior works on these three datasets ranges between 30\% and 62\%, as shown in Figure~\ref{fig:3}~(a). In Table~\ref{tab:first_main_table}, we compare \ourmethod\ with prior works on the Video Games dataset, where BIGRec achieves the second-best performance. The comparison between \ourmethod\ and BIGRec effectively contrasts two different approaches for finetuning LLMs in sequential recommendation tasks. Therefore, we conclude that the \ourmethod\ framework is more suitable than the finetune-and-retrieval framework for aligning LLMs with sequential recommendation problems. Figure~\ref{fig:3}~(b) shows that \ourmethod\ consistently outperforms BIGRec in NG@k and HR@k, and as $k$ increases, the percentage improvements become larger.

\paragraph{Inference Time}
Another advantage of \ourmethod\ is that it avoids generating multiple tokens during inference and instead provides recommendations based on the logits output. The inference speeds of BIGRec and \ourmethod\ are 0.066 and 6.5 samples per second respectively. \ourmethod\ achieves an almost 100$\times$ speed-up compared to the finetune-and-retrieval framework.

\begin{figure*}[!ht]
    \centering

    \subfloat[HR@5 comparison between CoVE and baseline methods on the Beauty, Sports, and Toys datasets.]{ \includegraphics[width=0.3\textwidth]{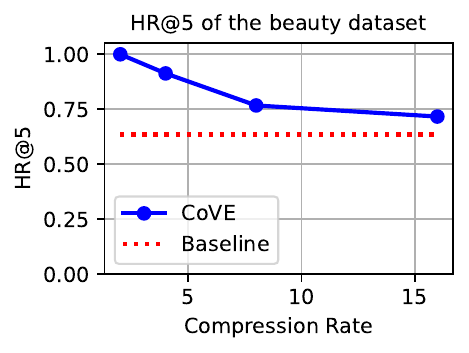}
   % \label{fig:beauty_hr5}
    \includegraphics[width=0.3\textwidth]{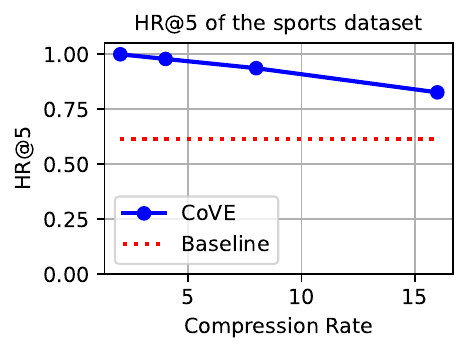}
   % \label{fig:sport_hr5} 
    \includegraphics[width=0.3\textwidth]{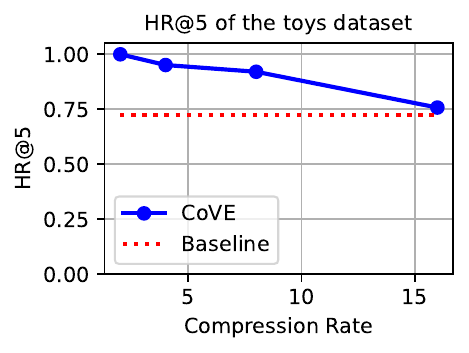}
   % \label{fig:toy_hr5}
}

    \subfloat[NG@5 comparison between CoVE and baseline methods on the Beauty, Sports, and Toys datasets.]{ \includegraphics[width=0.3\textwidth]{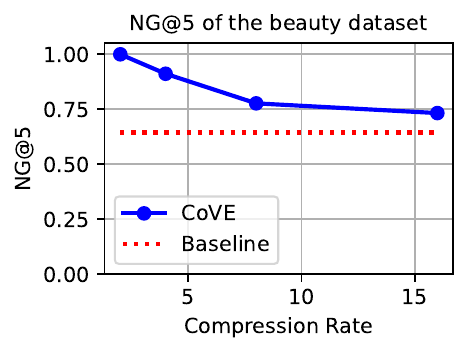}
   % \label{fig:ng_5}
    \includegraphics[width=0.3\textwidth]{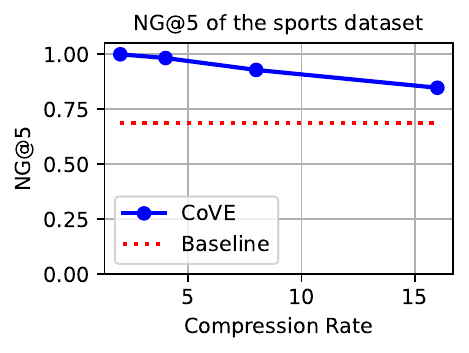}
   % \label{fig:sport_ng5} 
    \includegraphics[width=0.3\textwidth]{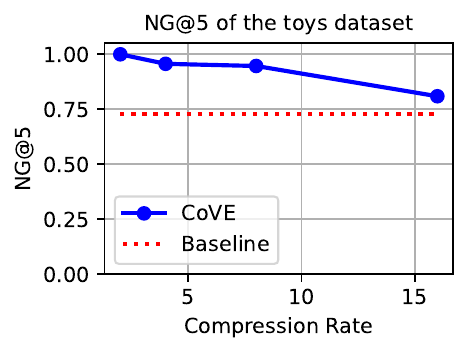}
   % \label{fig:toy_ng5}
}

\subfloat[HR@10 comparison between CoVE and baseline methods on the Beauty, Sports, and Toys datasets.]{ \includegraphics[width=0.3\textwidth]{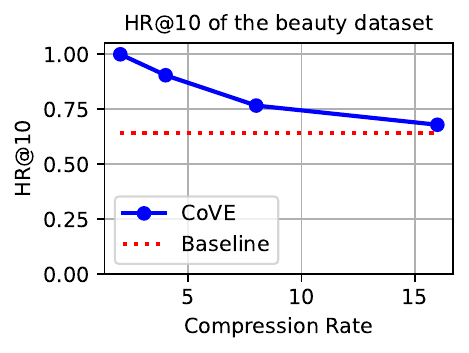}
    \includegraphics[width=0.3\textwidth]{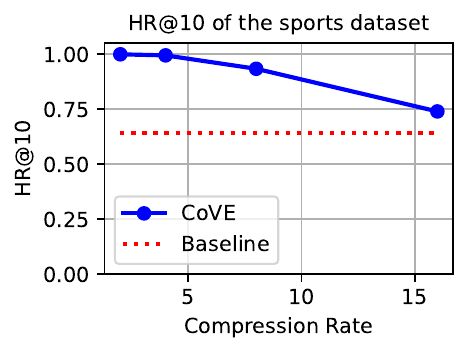}
    \includegraphics[width=0.3\textwidth]{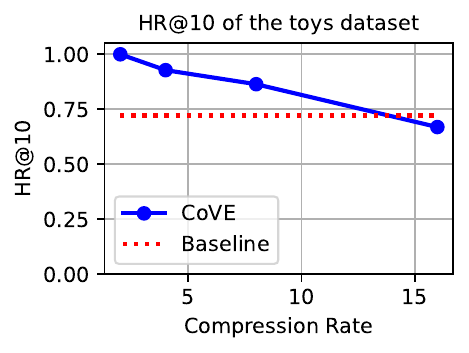}
}

    \subfloat[NG@10 comparison between CoVE and baseline methods on the Beauty, Sports, and Toys datasets.]{ \includegraphics[width=0.3\textwidth]{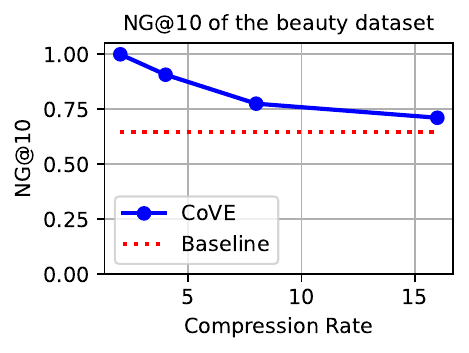}
    \includegraphics[width=0.3\textwidth]{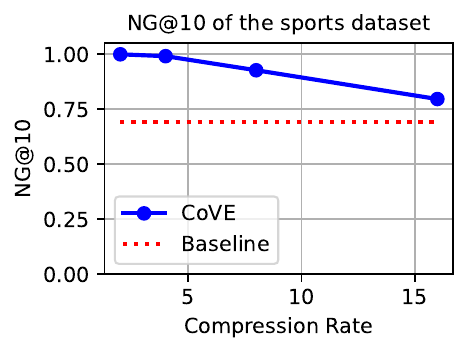}
    \includegraphics[width=0.3\textwidth]{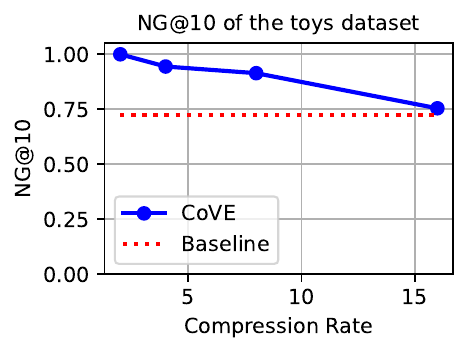}
}
    \caption{Normalized metrics, HR@5, NG@5, HR@10, and NG@10, of \ourmethod\ under different compression rates ($2, 4, 8, 16$), where the original embedding layer is compressed to $1/2, 1/4, 1/8,$ and $1/16$ of its original size, respectively. We compare four metrics across three datasets, resulting in twelve figures, showcasing \ourmethod\ alongside the state-of-the-art baseline. Notably, even when the embedding layer is compressed 16-fold, \ourmethod\ still outperforms the state-of-the-art baseline with only one exception (HR@10 of the toys dataset).}
    \label{fig:6figs}
\end{figure*}

\subsection{Ablation Study}

\begin{table}[!ht]
    \centering
    \begin{tabular}{p{0.9cm}p{0.5cm}p{0.95cm}p{0.95cm}p{0.95cm}p{0.95cm}}
        \toprule
        \textbf{Datas-ets} & {\bf I./E.} & \multirow{2}{*}{\textbf{NG5}} & \multirow{2}{*}{\textbf{NG10}} & \multirow{2}{*}{\textbf{HR5}} & \multirow{2}{*}{\textbf{HR10}} \\
        \midrule
        \multirow{3}{*}{\textbf{Beauty}} & {\bf E.}  & 0.045  & 0.0519 & 0.0622 & 0.0836 \\
        & {\bf I.} & 0.0057 & 0.0084 & 0.0094 & 0.0178  \\
        & {\bf Both} & 0.0498 & 0.0593 & 0.0714 & 0.1009 \\
        \midrule
        \multirow{3}{*}{\textbf{Toys}}   & {\bf E.} & 0.0429 & 0.0482 & 0.0566 & 0.0732 \\
        & {\bf I.} & 0.0058  & 0.0085 & 0.0095 & 0.018 \\
        & {\bf Both} & 0.0509 & 0.0595 & 0.0719 & 0.0986 \\
        \midrule
        \multirow{3}{*}{\textbf{Sports}} & {\bf E.} & 0.0213 & 0.0255 & 0.0308 & 0.0439 \\
        & {\bf I.} & 0.0053 & 0.0069 & 0.0079 & 0.0128 \\
        & {\bf Both}  & 0.0296 & 0.0359 & 0.0428 & 0.0624 \\
        \bottomrule
    \end{tabular}
    \caption{Performance metrics of \ourmethod\ under different settings. {\bf I.} represents using prompts with the item’s title information and setting the item embedding table as frozen. {\bf E.} represents using prompts without item's title information and setting the item embedding table as trainable. {\bf Both} represents using prompts with item's title information and setting the item embedding table as trainable, which is the default setting of \ourmethod. We present three datasets across four metrics.}
    \label{tab:7}
\end{table}

\paragraph{Importance of Item Title} 

Our prompt includes item titles as important textual information, as shown in Table~\ref{tab:input-output-form}. Here, we evaluate the importance of such text information in \ourmethod. For comparison, we finetune the same model with the same hyperparameters but use a different prompt format, where item titles are removed. Below is a tuning sample without items' titles:
\begin{itemize}
    \item \textbf{Task Instruction:} Given a list of video games the user has played, please recommend a new video game that the user likes to the user. 
    \item \textbf{Task Input:} Here is the list of video games that the user has played:
    <|16659|>, <|5641|>, <|4375|>, <|14747|>, where <|$\cdot$|> denotes the unique ID for each item in item space.
    \item \textbf{Task Output:} <|16673|>.
\end{itemize}

In Table~\ref{tab:7}, we present the evaluation metrics of \ourmethod\ with the two different prompts. We observe that \ourmethod\ without title information does not provide recommendations as effectively as \ourmethod\ with title information. However, \ourmethod\ does not completely fail in the absence of title information, which can be attributed to the strong sequence understanding ability of LLMs.

\paragraph{Importance of Tuning Embedding Table}

\ourmethod\ tunes both the item embedding table and transformer weights. We analyze the importance of tuning the embedding table by comparing the performance of \ourmethod\ with a trainable item embedding table to its performance with a frozen item embedding table. As in Table~\ref{tab:7}, the results indicate that \ourmethod\ with a frozen item embedding table performs very poorly. This suggests that learning high-quality embeddings for item IDs is crucial.

\paragraph{Different Embedding Layer Compression Rate}

Embedding layer compression plays a crucial role in \ourmethod, especially in large-scale industrial scenarios. Through a comparison of \ourmethod\ with different compression rates ranging from 2 to 16, we find that the performance of \ourmethod\ degrades as the compression rate increases, as shown in Figure~\ref{fig:6figs}. However, even with a compression rate of 16, \ourmethod\ still outperforms the baseline in HR@5 and NG@5.
We also observe that the robustness of \ourmethod\ to compression rates varies across datasets. For instance, \ourmethod\ remains highly robust to compression rates up to 8 on the Sports \& Outdoors and Toys \& Games datasets. However, \ourmethod\ is more sensitive to higher compression rates on the Beauty dataset, where normalized NG@5 and HR@5 degrade to 0.75 at a compression rate of 8.

\begin{figure}[!ht]
  \includegraphics[width= 0.45\textwidth]{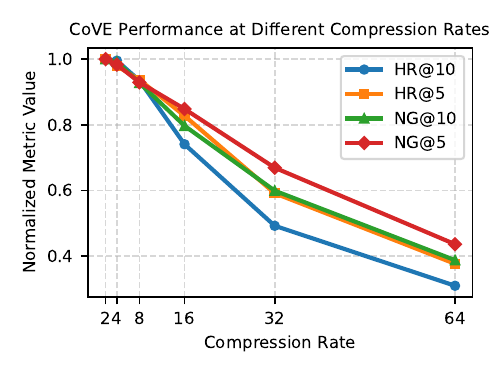}
  \caption{Performance of \ourmethod\ on the Beauty dataset under increasing embedding compression rates (2$\times$ to 64$\times$). 
  }
  \label{fig:beyond_8}
\end{figure}

To visualize the performance of \ourmethod\ with higher compression rate, we provide performance of \ourmethod\ on the Beauty dataset with compression beyond 16, as shown in Figure~\ref{fig:beyond_8}.

\section{A Case Study}

In this section, we analyze what an LLM can learn after finetuning under the \ourmethod\ framework. To show this, we give an example of the input and the generated output of finetuned LLM by \ourmethod:
\begin{itemize}
    \item \textbf{Input:} Instruction: Given a list of items the user has bought before, please recommend a new item that the user likes to the user. Here is the list of items that the user has bought: <|1|>: ``Hasbro Electronic Catch Phrase'', <|2|>: ``Gloom'', <|3|>: ``Cards Against Humanity'', <|4|>:  ``Carcassonne Basic Game'', <|5|>: ``Asmodee 7 Wonders Wonder Pack''.
    \item \textbf{Generated Output:} <|5|>: ``Asmodee 7 Wonders Wonder Pack'', <|1441|>: ``Cards Against Humanity: First Expansion''.
\end{itemize}
The generated content strictly follows the format of the user's interaction history, i.e., each item's ID is followed by its title. We find that the finetuned LLM consistently outputs the correct item title corresponding to its ID. This indicates that under the \ourmethod\ framework, the finetuned LLM learns the mapping between an item's ID and its title.
We argue that this is crucial for high-quality recommendations. One key advantage of \ourmethod\ is its ability to convert the item title prediction task into an item ID prediction task. While the latter is easier than the former—since the number of tokens to predict is significantly smaller—item ID prediction is meaningful only when the LLM correctly understands the relationship between IDs and titles.

\section{Related Work}

\paragraph{LLMs in Sequential Recommendation}

Currently, LLMs are primarily utilized in recommendation systems in two ways: providing embeddings for items \cite{zhang2021unbert, wu2021empowering, liu2021pre} and directly generating recommendations \cite{bao2023bi, bao2023tallrec, hegselmann2023tabllm, li2023gpt4rec, ngo2024recgpt, zhang2024finerec}.

Our work is more related to generative recommendation systems.
For instance, \citet{bao2023tallrec} proposes a finetuning framework to align LLMs with recommendations, where the task is defined as a Yes-No question-answering problem. However, under this framework, generating the next item for users is challenging. GPT4Rec \cite{li2023gpt4rec} integrates LLMs with a search engine, where the LLM generates multiple queries representing the user's interests, and the search engine outputs the final recommendation.
The existing work most similar to our approach is the finetune-retrieval framwork adopted in \cite{bao2023bi, bzk+24}. During training, BIGRec finetunes an LLM on a recommendation dataset. During inference, BIGRec computes the distances between the embedding of the sentence generated by the finetuned LLM and the embeddings of all items in the item space. Based on these distances, BIGRec determines the next recommended item for the user. Our approach is similar to BIGRec in the training stage; however, in the inference stage, we do not need to generate content using LLMs. Instead, we provide recommendations directly based on logits of items. This approach reduces inference time and eliminates hallucinations. The trade-off is a larger embedding layer, which we mitigate through embedding layer compression. \citet{liu2025llmemb} studies the long-tail problem in sequential recommendation by enhancing item embeddings for rare items. It uses a pre-trained LLM (LLaMA-7B) and adapts it to the recommendation domain via contrastive fine-tuning (SCFT) and recommendation adaptation, and focuses on preserving semantic richness while integrating collaborative signals.

\paragraph{Vocabulary Expansion of LLMs}
Our study extends the vocabulary of pretrained LLMs to assign each item a unique embedding, thereby enhancing the performance of LLM-based recommendation systems. This technique is closely related to vocabulary expansion in domain-specific LLMs.
Pretrained LLMs are typically trained on a large corpus that differs in distribution from domain-specific corpora. Thus, vocabulary expansion has become a key strategy to adapt LLMs pretrained on general datasets to domain-specific applications, improving decoding efficiency and semantic comprehension ability \cite{liu2024gold, kajiura2023vocabulary}.

For instance, \citet{cui2023efficient} continue training LLaMA models on a Chinese corpus with additional tokens, leading to improved proficiency in understanding and generating Chinese content. \citet{liu2023task} propose a task-specific tokenization approach to adapt the generation pipeline for the mental health domain. Similarly, \citet{liu2024gold} introduce an adaptive method for vocabulary expansion in domain-specific LLMs. \citet{liu2023chipnemo} train a new tokenizer on domain-specific data and expand the existing general-purpose tokenizer by incorporating newly identified tokens to adapt LLMs for chip design.

\paragraph{Embedding Layer Compression}

One branch of the embedding layer compression technique is low-precision methods which focus on reducing the number of bits used to represent each weight in the embedding table. Due to variations in bit width and their unique advantages, low-precision methods can be further categorized into binarization and quantization.

Binarization methods compress full-precision weights into binary codes (1-bit representation) \cite{42,77,33,76,37}. Other notable works include discrete collaborative filtering (DCF) \cite{73}, which learns binary embeddings for collaborative filtering; candidate generation and re-ranking (CIGAR) \cite{28}, which utilizes the scaled $\tanh$ function for binary approximation; hashing with graph neural networks (HashGNN) \cite{54}; and low-loss quantized graph convolutional networks (L$^2$Q-GCN) \cite{5}, which propose variations using straight-through estimators for gradient computation during training.

Quantization methods map 32-bit floating-point weights to lower-precision representations using multiple bits rather than just one \cite{66}. Moreover, \citet{16,69} study both uniform and non-uniform quantization techniques. To better preserve model accuracy, \citet{31} proposes adaptive low-precision training (ALPT), which adaptively adjusts quantization step sizes during training.

\paragraph{Generating Item Embeddings}

The use of LLMs in recommendation systems marks a significant shift from traditional embedding approaches. While earlier techniques primarily relied on collaborative filtering \cite{sfhs07} or simple word embeddings \cite{msd+15}, LLMs offer a more sophisticated understanding of items through their pre-trained knowledge and contextual processing capabilities.

Several studies have demonstrated the effectiveness of LLM-generated embeddings in capturing complex item characteristics and relationships. For instance, \citet{zhang2021unbert} and \citet{wu2021empowering} both examine the application of pre-trained language models to news recommendations. Specifically, \citet{zhang2021unbert} propose a BERT-based user-news matching model that not only leverages the extensive language knowledge of the pre-trained model to enhance textual representation but also captures multi-grained user-news matching signals at both the word and news levels. In contrast, \citet{wu2021empowering} explore different variations of the BERT model.
Additionally, \citet{liu2021pre} investigate a recent state-of-the-art Chinese pre-trained language model called Enhanced Representation through Knowledge Integration (ERNIE).

\section{Conclusion}
We propose a novel framework, \ourmethod, to better align LLMs with sequential recommendation problems. The core idea of \ourmethod\ is to assign each item a unique ID that can be encoded as a token and given a unique embedding. The item embedding table is finetuned along with the weights in transformer blocks during training.
\ourmethod\ outperforms baselines by up to 62\% in NG@5, NG@10, HR@5, and HR@10 on Amazon review datasets. Additionally, \ourmethod\ achieves approximately 100$\times$ speed-up in inference time compared to the existing finetune-and-retrieval framework.
Through comprehensive ablation experiments, we find that including an item's textual information enhances \ourmethod's performance. From case studies, we observe that \ourmethod\ effectively memorizes ID-title pairs for different items, which is crucial for providing high-quality recommendations. In this paper, we used AI tools solely for language assistance.

\section*{Limitations}
Embedding layer compression is an important part of \ourmethod. In this study, we adopt hash function to compress embedding layers. There are more advanced compression method that can be further explored. Besides, from the memory efficiency perspective, more works can be done on the robustness of \ourmethod \ to other memory-efficient methods, such as memory-efficient optimizers, quantization, low-rank approximation, etc.

\bibliography{custom}

\appendix

\section{Description of Prior Methods} \label{app:prior method}
GRU4Rec \cite{gru4rec} applies a GRU-based RNN for sequential recommendations. BERT4Rec \cite{sun2019bert4rec} is a bidirectional self-attention model for the sequential recommendation problem. Caser \cite{caser} embeds recent items in the iteration sessions into latent spaces and uses convolutional filters to learn sequential patterns. SASRec \cite{sasrec} proposes a self-attention-based sequential model to balance the advantage of Markov Chains and the advantage of RNN. P5 \cite{p5} converts various data formats into natural language and builds on pretrained T5 to learn sequential interaction patterns. HGN \cite{hgn} adopts gating modules to control the flow of item features and learn item relations. S$^3$-Rec \cite{s3rec} applies self-supervised learning to learn correlations among data. FDSA \cite{fdsa} integrates different features of items into feature sequences and then models item transition patterns and feature transition patterns. TIGER \cite{tiger} first learns semantic IDs of different items and then trains a generative recommendation system based on the semantic IDs. DROS \cite{dros} is a framework to enhance the generalization of recommendation systems that is more suitable for streaming data. GPT4Rec \cite{li2023gpt4rec} introduces a novel generative framework for personalized recommendation systems that addresses several limitations of existing NLP-based recommender systems, such as treating items as only IDs and using discriminative modeling. BIGRec \cite{bao2023bi} uses a finetune-and-retrieval framework to align LLM with sequential recommendation tasks.

\section{More Related Work}

\subsection{Embedding Layer Compression in Recommendation Systems}

Embedding layer compression techniques for recommender systems can be broadly categorized into three main areas, low-precision methods, mixed-dimension methods, and weight-sharing methods \cite{lgt+24}. Each category addresses different aspects of reducing memory usage while maintaining model performance.

\paragraph{Mixed Dimension}

Second, mixed-dimension techniques optimize memory usage by assigning different embedding dimensions to different features. Similarly, due to the different characteristics that prior works focus on, mixed-dimension techniques can be further categorized into rule-based approaches, neural architecture search (NAS) based approaches, and pruning.

Rule-based approaches, such as those proposed by \citet{15,53}, determine embedding dimensions using predefined heuristics based on feature frequency or field size. The advantage of rule-based approaches is their computational efficiency; however, they may not achieve optimal compression. For example, the compressed sequential recommendation framework (CpRec) proposed by \citet{53} divides a group of items into subgroups, each of which is assigned a predefined dimension based on feature frequencies. Similarly, \citet{15} assigns a predefined dimension based on the number of features included in a particular field.

NAS-based approaches automatically determine optimal embedding dimensions. Unlike rule-based approaches, which use predefined dimensions, NAS-based approaches learn the optimal embedding dimensions \cite{39,59,79}. To achieve this, \citet{25} present neural input search (NIS), which utilizes reinforcement learning; \citet{78} analyze automated embedding dimensionality search (AutoEmb), which employs differentiable architecture search; and \citet{3,43} respectively propose recommendation with universally learned elastic embeddings (RULE) and the optimal embedding table learning framework (OptEmbed), both of which leverage evolutionary search strategies.

Pruning techniques optimize memory usage by identifying and removing less important weights, thereby creating sparse embedding tables \cite{41,48}. \citet{10} use iterative pruning with retraining, whereas \citet{7,68} employ learnable masks to determine which weights to retain.

\paragraph{Weight Sharing}

Weight-sharing approaches reduce memory usage by allowing multiple features to share embedding parameters, which can be categorized into two sub-types: hashing and vector quantization.

Hashing functions map features to shared embeddings \cite{47, 72, 12}. For example, \citet{50} uses multiple hash functions to generate two index vectors and maintain two meta-tables. Furthermore, \cite{67} develops binary code based hash (BCH) that operates features at the bit level. Random offset block embedding (ROBE) \cite{11} uses hash functions so that all elements of the embedding table can be mapped to a shared memory.

Vector quantization methods cluster similar embeddings together \cite{34}. Similarity-aware embedding compression (Saec) \cite{62} uses traditional clustering, while subsequent works like multi-granular quantized embeddings (MGQE) \cite{26} and extremely memory-efficient factorization machine (xLightFM) \cite{24} employ product quantization. Linear-time self attention (LISA) \cite{63} specifically addresses quantization for self-attention mechanisms in recommender systems.

\subsection{Recommender System and LLM at a Broader Scale}

To ensure efficient training or deploying \ourmethod\ at scale, we also need LLM optimization to reduce memory or computation \cite{zyw25,lsx+25,gswy25,gsy25,sxy25,lls+24,gsy23,syz24,swy23,lswy23,cll+25_mamba_tc0,chl+24_rope,cll+25_universality,cll+25_kvcache}. In particular, \cite{zyw25} proposes the selection of subspaces of importance sampling (I3S) for low-rank optimization to enable memory-efficient training of LLMs. \cite{gswy25,gsy25,gsy23,syz24,swy23,lswy23} analyze the attention inspired regression problem to reduce the computational complexity. \cite{sxy25} builds upon a linear time variant of the softmax attention problem. \cite{lsx+25} analyzes the dynamic maintenance of kernel density estimation. \cite{lls+24} develops the convolution basis to reduce the computational cost of masked attention. \cite{chl+24_rope,cll+25_universality} analyze the efficiency of RoPE attention and visual autoregressive transformers. \cite{cll+25_mamba_tc0} makes use of the concept of circuit complexity to study the computational limitation of Mamba.

Additionally, recommender systems also exhibit a strong connection with reinforcement learning, as both aim to model sequential decision-making processes where the system learns optimal strategies (e.g., item recommendations) by interacting with users and observing feedback over time \cite{zcy23,zcz25}. From the perspective of LLMs, our work can be inspirational for and related to other possible application where LLMs are not very closely involved yet \cite{zpt+22,zlp+23}.   
\end{document}